\documentclass{article}

\usepackage{PRIMEarxiv}

\usepackage[utf8]{inputenc} % allow utf-8 input
\usepackage[T1]{fontenc}    % use 8-bit T1 fonts
\usepackage{url}            % simple URL typesetting
\usepackage{booktabs}       % professional-quality tables
\usepackage{natbib}
\usepackage{amsfonts}       % blackboard math symbols
\usepackage{nicefrac}       % compact symbols for 1/2, etc.
\usepackage{microtype}      % microtypography
\usepackage{lipsum}
\usepackage{fancyhdr}       % header
\usepackage{graphicx}       % graphics
\graphicspath{{media/}}     % organize your images and other figures under media/ folder

\usepackage{amsmath}

\usepackage{epstopdf}
\usepackage[title]{appendix}
\usepackage{flushend}

\usepackage{subfigure}
\usepackage[colorlinks=true, allcolors=blue]{hyperref}
\usepackage[table,xcdraw]{xcolor}
\usepackage{textcomp}
\usepackage[figuresright]{rotating}
\usepackage{tablefootnote}
\usepackage[justification=centering]{caption}
\usepackage{xcolor}

\title{Identifying Experts in Question \& Answer Portals: \\ A Case Study on Data Science Competencies in Reddit
\thanks{
\textcolor{red}{This is a pre-print version of the article.}} 
}

\author{
  Sofia Strukova, Jos\'e A. Ruip\'erez-Valiente, F\'elix G\'omez M\'armol \\
  Department of Information and Communications Engineering \\
  University of Murcia \\
  Murcia (Spain)\\
  \texttt{\{strukovas, jruiperez, felixgm\}@um.es} \\
}

\usepackage{amssymb}
\usepackage{subfigure}
\usepackage{color}
\usepackage{pifont}

\definecolor{ao}{rgb}{0.0, 0.5, 0.0}
\newcommand{\yestick}{{\color{ao}\ding{51}}}
\newcommand{\notick}{{\color{red}\ding{55}}}

\begin{document}
\maketitle

\begin{abstract}
The irreplaceable key to the triumph of Question \& Answer (Q\&A) platforms is their users providing high-quality answers to the challenging questions posted across various topics of interest. From more than a decade, the expert finding problem attracted much attention in information retrieval research. Based on the encountered gaps in the expert identification across several Q\&A portals, we inspect the feasibility of identifying data science experts in Reddit. Our method is based on the manual coding results where two data science experts labelled not only expert and non-expert comments, but also out-of-scope comments, which is a novel contribution to the literature, enabling the identification of more groups of comments across web portals. We present a semi-supervised approach which combines 1,113 labelled comments with 100,226 unlabelled comments during training. The proposed model uses the activity behaviour of every user, including Natural Language Processing (NLP), crowdsourced and user feature sets. We conclude that the NLP and user feature sets contribute the most to the better identification of these three classes. It means that this method can generalise well within the domain. Finally, we make a novel contribution by presenting different types of users in Reddit, which opens many future research directions.
\end{abstract}

% keywords can be removed

\section{Introduction}
\label{sec:introduction}

Learning is no longer purely based on knowledge transference from teacher to student. Instead, many people are learning on their own, either by self-regulating with courses or books or by interacting with communities of users online. Consequently, there is not only much literature focused on traditional learning and its new trends~\citep{razeeth2019learning} but also much work done on exploring informal learning happening across digital environments that are not targeting to acquire new knowledge or competencies~\citep{strukova:inpress}. For example, in Question \& Answer (Q\&A) portals, millions of daily active users worldwide~\citep{aabdelaziz:forums} discuss a great variety of knowledge domains, create new content, refer to the existing one, and participate in constructive discussions. Users are learning across these portals and through the actions mentioned above. Moreover, new skills are spanning across a great variety of fields, not being limited only to common conversation topics. This is because there are many sites concentrated on questions helpful for different careers.

Q\&A portals are being built based on the data that users are generating on a daily basis~\citep{ansari2020identifying}. Users are asking many questions across these community platforms, even though the World Wide Web already contains millions of answers to most questions in one way or another. Yet, users struggle to find the answers to their questions in a single portal or do not have the capability to formulate the question as a keyword-based search~\citep{graham2019society}. As might be expected, some users make more relevant contributions to the community than others. Consequently, relevant domain-specific contributions are often mixed with unhelpful content. Therefore, we identify the need to improve the detection of expert users based on the evidence of these portals, as it can have multiple potential applications such as a better matchmaking of experts to questions, support in formal education, or self-awareness reflection.

By estimating the user expertise, we can infer the quality of content because users of higher expertise tend to produce higher quality content~\citep{lim2017estimating}. Moreover, through the estimation of user expertise of content authors, it is possible to infer the information quality of the content despite the lack of user votes. Additionally, social networks are a primary source of news and information that can be steered, distorted, and influenced~\citep{zago2019screening}. In this way, we could detect potential malicious or unreliable users and reduce their influence.

In this work, we examine several Q\&A portals and the existing studies focusing on expert identification. We selected Reddit as a Q\&A portal for our study because its full potential in expert finding is still uncovered. The encountered gaps of the explored studies are associated with the deployed type of features and the fact that the maximum number of categories that the authors of these articles analysed is two -- expert and non-expert. Based on these grounds, we propose our method to detect not only expert and non-expert comments, but also out-of-scope comments in Reddit, which is a novel contribution to the literature, enabling the identification of more groups of comments across web portals. Moreover, a dominant contribution of this work is a description of a detailed procedure for manual labelling of these three groups of comments that can be replicated by other work. At last, as far as we are aware, this is the first time that a characterisation of expert, non-expert and out-of-scope users is presented in any Q\&A portal. Accordingly, we state the following Research Questions (RQs) for our work:

\begin{itemize}
    
    \item \textbf{RQ1}. Is it feasible to distinguish between expert, non-expert and out-of-scope comments across Reddit by analysing comments of the respective thread(s)?
    
    \item \textbf{RQ2}. What is the most important feature set for the detection of topical expertise among Natural Language Processing (NLP), crowdsourced, user-author feature sets and their combinations?
    
    \item \textbf{RQ3}. What are the common characteristics of the expert, non-expert and out-of-scope comments?
    
    %\item \textbf{RQ}. What is the impact of using two classes (expert, non-expert) over three classes (expert, non-expert, out-of-scope)?
    
    \item \textbf{RQ4}. Can we find different profiles of users in Reddit, including expert users, non-experts and out-of-scope users?
    
\end{itemize}

The remainder of this paper is structured as follows. In Section~\ref{sec:background}, we focus on the background of our study covering the subject of Q\&A portals, topical experts identification and corresponding methods. In Section~\ref{sec:methodology}, we present our research methodology. We expand this section by selecting the Q\&A portal, the thread for expert finding and their characteristics. Next, we depict the human-in-the-loop process and the final data collection. Then, we describe the ML models selected for solving the task of expert finding and the feature engineering process. Our findings are outlined in Section~\ref{sec:results}, while we extend the results in Section~\ref{sec:discussion}. Finally, we draw our conclusions and future research directions in Section~\ref{sec:conclusion}.

\section{Background}
\label{sec:background}

There is a large body of literature examining Q\&A portals. Three perspectives were essential for this study, which we outline in this section. First, we discuss the most commonly used portals and their characteristics. Next, we review the issue of topical experts and suitable Q\&A portals for identifying them. Finally, we explore the methods used for identifying experts across selected platforms.

\subsection{Question \& Answer Portals}

The main goal of Q\&A platforms is to provide users with an opportunity to post messages asking what they are interested in and/or replying to others. The main elements of the majority of Q\&A portals are users registered in the system, questions asked by users, answers provided by fellow users and opinions expressed in the form of votes and comments~\citep{gyongyi2007questioning}. This section highlights the use and differences between some of the most popular Q\&A portals.

Answers.com\footnote{\url{https://www.answers.com/}} formerly known as WikiAnswers, is an Internet-based knowledge exchange platform mostly for students of all ages. Over and above that, Reddit\footnote{\url{https://www.reddit.com/}} is famous for its sense of community, having a big variety of `subreddits,' which can be defined as specific communities dedicated to particular topics for any and every interest, and if there is not, it can be easily created. By way of an alternative, Stack Overflow\footnote{\url{https://stackoverflow.com/}} is a portal that developers frequently use to post their programming questions and where fellow developers provide answers~\citep{nasehi2012makes}. In turn, Quora\footnote{\url{https://www.quora.com}} differs from the above-mentioned portals by integrating a social network into its basic structure~\citep{wang2013wisdom}. Unlike the rest of the websites, AskFm\footnote{\url{https://ask.fm/}} is a social networking platform that uses a Q\&A format to connect users with each other through conversational exchanges~\citep{farrugia2019have}. Finally, we also explored
Yahoo! Answers\footnote{\url{https://answers.yahoo.com/}} despite the fact that it was closed in 2021\footnote{\url{https://help.yahoo.com/kb/SLN35642.html}}, since it was one of the leading portals allowing users to make contributions to a topic of interest, receive assistance from others, and interact with each other in the community~\citep{zhang2020public}.

In Table~\ref{tab:qa_comparison} we present a comparison across several characteristics of the leading Q\&A portals, namely, Answers.com, AskFm, Quora, Reddit, Stack Overflow and Yahoo! Answers. It is worth noting that there are other Q\&A portals that were not included in this comparison but due to the space limits we selected the most prominent ones. All the portals presented in Table~\ref{tab:qa_comparison} are suitable places for seeking advice, gathering opinions, and satisfying the curiosity about a countless number of things. These Q\&A portals can cover all sorts of questions -- from mundane and everyday questions to complex and expert ones~\citep{adamic2008knowledge}. We observed the fact that due to a large amount of traffic across these portals, questions often receive multiple answers within a short period of time but also quickly disappear in the flood of new questions. AskFm is the only social portal where users connect with each other through the conversational exchange. Answers.com is an encyclopedia-like portal with knowledge from reputable sources where anyone can edit the answer or question, meaning that there is only one answer that several people have written. Contrastingly, Yahoo! Answers has one question with multiple answers, so do Stack Overflow, Quora and Reddit. All the portals operate worldwide and require registration in order to contribute, while there is no need to authenticate to browse. 

\begin{table*}[]
\resizebox{\textwidth}{!}{%
\begin{tabular}{|l|l|l|l|l|l|l|l|l|}
\hline
\textbf{\begin{tabular}[c]{@{}l@{}}Question \& \\ Answer portal\end{tabular}} & \textbf{\begin{tabular}[c]{@{}l@{}}Foundation\\ year\end{tabular}} & \textbf{\begin{tabular}[c]{@{}l@{}}Area \\ served\end{tabular}} & \textbf{\begin{tabular}[c]{@{}l@{}}Languages \\ available\end{tabular}} & \textbf{\begin{tabular}[c]{@{}l@{}}Registration \\ to browse/\\ contribute\end{tabular}} & \textbf{\begin{tabular}[c]{@{}l@{}}Monthly \\ active users\end{tabular}} & \textbf{Focus} & \textbf{\begin{tabular}[c]{@{}l@{}}User can mark an \\ answer as correct\end{tabular}} & \textbf{API} \\ \hline
Answers.com & 2005 & Worldwide & \begin{tabular}[c]{@{}l@{}}6 languages\\ incl. English\end{tabular} & \notick/\yestick & - & General & \notick & \notick \\ \hline
AskFm & 2010 & Worldwide & \begin{tabular}[c]{@{}l@{}}49 languages\\ incl. English\end{tabular} & \notick/\yestick & 13 million & \begin{tabular}[c]{@{}l@{}}Social\\ topics\end{tabular} & - & \notick \\ \hline
Quora & 2009 & Worldwide & \begin{tabular}[c]{@{}l@{}}24 languages\\ incl. English\end{tabular} & \notick/\yestick & 150 million & General & \notick & \notick \\ \hline
Reddit & 2005 & Worldwide & \begin{tabular}[c]{@{}l@{}}8 languages\\ incl. English\end{tabular} & \notick/\yestick & 430 million & General & \notick & \yestick \\ \hline
StackOverflow & 2008 & Worldwide & \begin{tabular}[c]{@{}l@{}}5 languages\\ incl. English\end{tabular} & \notick/\yestick & 50 million & \begin{tabular}[c]{@{}l@{}}Computer\\ Science\end{tabular} & \yestick & \yestick \\ \hline
Yahoo!Answers & 2005 & - & \begin{tabular}[c]{@{}l@{}}12 languages\\ incl. English\end{tabular} & \notick/\yestick & - & General & \yestick & \yestick \\ \hline
\end{tabular}%
}
\caption{Question \& Answer Portals Comparison}
\label{tab:qa_comparison}
\end{table*}

There were three main points of comparison for our research: the amount of monthly active users, the availability of the feature to mark the answer as correct, and the access to an Application Programming Interface (API). We were not able to find the number of active users per month in Answers.com (not available on the official webpage) and Yahoo! Answers (it ceased operations in most languages on May 4, 2021), while across others, the most visited portal is Reddit with its 430 million active users per month, followed by Quora and Stack Overflow with 150 and 50 million monthly active users, respectively. Only on Stack Overflow and Yahoo! Answers users are/were able to mark an answer as accepted or the best. Finally, half of the portals that we explored offer an API -- Reddit, Stack Overflow and Yahoo! Answers.

\subsection{Topical experts and suitable platforms for their finding}

The portals above sustain large amount of traffic, and the posted questions often receive multiple answers within a short period of time. It brings the importance of quickly evaluating the answers according to the help they provide in general or specific topics. This would be beneficial for new users who are still not familiar with the system and community. Alternatively, expert identification can be used for an expert recommendation service in a social Q\&A site. Unfortunately, little is known about the properties of experts and non-experts and how to detect experts in general or in specific topics~\citep{patil2016detecting}.

Many questions are meant to trigger discussions or to encourage users to express their opinions. Sometimes, new questions in these Q\&A portals stimulate respondents to disseminate curated knowledge that may not be available on other websites, or it may take time for a user to find, understand and summarise relevant information from other sites. However, in most portals, users can evaluate answers by upvotes and downvotes. Via these interactions, it is feasible to naturally reveal the best answer for a question.
 
The portals that we explored earlier offer different metrics to identify more reliable users. For example, Quora chooses a small subset of registered users to be reviewers and have the power to flag or remove low-quality answers and questions~\citep{wang2013wisdom}. Contrastingly, in Reddit, users themselves build the community by earning or losing “karma points” when their content or comments receive upvotes or downvotes, respectively. This reward system encourages users to post good content, make valuable comments, and provide relevant feedback~\citep{anderson2015ask}. Although the presence of karma and awards gives the Reddit community the ability to assess the relevance of specific users whose karma suggests that they are more highly valued within the collective, high karma does not mean that the user has the experience and/or knowledge of the field. Contrastingly, Stack Overflow and Yahoo! Answers provided the ability to the questioner to choose the best answer, which could be the one that agrees with the questioners’ opinions, while for entertainment categories, the wittiest reply may win~\citep{adamic2008knowledge}. Accordingly, we can conclude that the answers selected as the best are very subjective. Moreover, the majority of users across these portals are questioners (who are likely non-experts), which could be partly explained by the fact that these are places to “ask questions” ~\citep{gyongyi2007questioning}. Therefore, we again emphasise the importance of identifying experts who are able to provide factual answers. 

\subsection{Methods for expert finding}

There are many studies focusing on the identification of topical experts. In the scope of our work, we are mainly interested in the methods applied for solving this task. Following the comparison in Table~\ref{tab:qa_comparison} presented in the previous section, we explore how the experts were found in each of these platforms. We did not find any work aiming to identify experts in the AskFm and Answers.com forums. Therefore, this section explores the most representative methods to identify experts across the rest of the platforms, namely, Quora, Reddit, Stack Overflow and Yahoo! Answers. It is pertinent to note that there are studies focused on the identification of topical experts in other platforms, e.g., Qian et al. explored Stack Exchange platform \footnote{\url{https://stackexchange.com/}} which is a network of sites that releases data from various Q\&A portals~\cite{QIAN2022107842}. In turn, we decided to explore its most visited community Stack Overflow.
%It can be explained by the fact that AskFm is a social website so, a priori, there is no correct answer because most of the questions are personal. On the other hand, questions in Answers.com are oriented on helping with homework so, typically, there are not many answers that are often correct since users do not answer if they are unsure. In general, questions are straightforward and do not suggest a possibility of lengthy discussion.

\subsubsection{Quora}

The most demonstrative example of identifying experts in Quora is shown by Patil et al., who performed an analysis of the behaviour of experts and non-experts in five popular topics~\citep{patil2016detecting}. After manually labelling the Quora dataset to get the ground truth (i.e., which profile is an expert’s profile or a non-expert’s profile) and analysing behaviours of experts and non-experts, the authors concluded that their activities and linguistic characteristics are different. Based on these observations, Patil et al. proposed three groups of features (activity features, quality of answer features and linguistic features) and developed topic-specific classifiers based on these features to prove that it is feasible to detect experts in each topic-based dataset.

\subsubsection{Reddit}

Kassing et al. investigated relevant domain-specific content in Reddit in the form of submissions shared by experts and non-experts~\citep{kassing2015locating}. Their results revealed that the overall number of identified knowledgeable users (more than one hundred thousand) and relevant domain-specific contents qualify Reddit as a platform for expert finding. However, there are not many studies focused on this topic across this portal. One of the existing ones focused on the estimation of Reddit user expertise in the form of the information quality of user-generated content by predicting the users’ contribution quality through their estimated expertise instead of focusing on the content itself~\citep{lim2017estimating}. The concrete metrics that were used are as follows: 1) Contribution Count which makes the assumption that experts are those users who make a large number of significant contributions as judged by the vote difference and the number of good contributions made by the users as their estimated user expertise; 2) Contribution Z-Index considers how many times a user has made a good contribution rather than a bad one; 3) Contribution Score which measures the significance of user contributions as a score by counting the vote difference of user contributions, where experts are users with high collected scores from their comments; and 4) Contribution Rating which is a comparison approach for user interactions where each thread is modelled as a competition between the users who comment on it and the performance of the users is measured given the significance of their contribution (a user of higher rating is expected to be more likely to contribute significant contributions). Finally, the authors applied these metrics to predict the information quality of user-generated content, concluding that it is possible to estimate user expertise for the prediction of content quality.

On the other hand, Choi et al. focused on conversational patterns in terms of users’ volume, responsiveness, and virality applying NLP techniques. First, the authors performed a semantic analysis using text analysis software called Linguistic Inquiry and Word Count (LIWC) that counts words belonging to psychologically meaningful categories. Next, the authors measured if the readability difficulties of titles and texts are related to their volume, responsiveness, and virality. In the final step, Choi et al. computed the message relevance by measuring each word’s term frequency-inverse document frequency (TF-IDF). They found that extensive, responsive, and viral conversations tend to have high document relevancy between parent and child comments and are likely to have complex texts, whereas a responsive conversation tends to have plain texts~\citep{choi2015characterizing}.

\subsubsection{Stack Overflow}

Across studies focusing on finding experts on Stack Overflow, we found various methods applied to solve this task. The authors of~\citep{van2015early} extracted textual, behavioural, and time-aware features in order to build a semi-supervised machine learning (ML) approach to predict whether a user will become an expert in the long term.

In contrast, Faisal et al. proposed statistical expert-ranking techniques including 1) Exp-PC which is an adaptation of g-index~\citep{egghe2006theory} (an author-level metric which measures the bibliometric impact of researchers) for ranking experts on Stack Overflow forum; 2) Rep-FS taking into account voters reputation and upvote ratio; and 3) Weighted Exp-PC which computes user expertise by combining their Exp-PC and Rep-FS scores. Respectively, the authors measured users’ reputation and expertise given both the quality of their answers and their consistency in providing quality answers. They explained the choice of the methods by the fact that most existing expert-ranking approaches consider basic features, such as the total number of answers provided by a user, but ignore the quality and consistency of the user’s answer~\citep{faisal2019expert}.

Contrariwise, Riahi et al. explored the viability of NLP methods such as 1) TF-IDF reflecting how important a word is to a document in a collection or corpus; 2) Language Model, which considers that rare terms in the corpus occur in only a group of documents in the corpus and have a significant influence on the ranking; 3) Latent Dirichlet Allocation (LDA) which is a three-level hierarchical Bayesian model informative about the content of users’ profiles where each profile is composed of questions containing sentences~\citep{riahi2012finding}.

\subsubsection{Yahoo! Answers}

Adamic et al. found Yahoo! Answers to be an astonishingly active social world with a great diversity of knowledge and opinion being exchanged~\citep{adamic2008knowledge}. The authors stated that the knowledge that users were sharing in this portal was very broad but generally not very deep. More than that, they found that some categories of Yahoo! Answers resembled a technical expertise sharing forum, while others had different dynamics in terms of support, advice, or discussion. 

All the research centred on Yahoo! Answers forum and identifying experts across it mainly applied network analysis methods. The most representative work of this platform is~\citep{bouguessa2008identifying}. Bouguessa et al. stated that most existing approaches attempting to discover experts model the environment as a graph in which the nodes represent users and the edges represent their interactions. In this way, the authors explored the most effective and widely used link analysis techniques, namely, PageRank, Hyperlink-Induced Topic Search (HITS), Z-score and InDegree algorithm. For example, according to the HITS algorithm, the quality of a page as an authority depends on the quality of the pages that point to it as hubs and vice versa. After examining those algorithms, Bouguessa et al. proposed to model the expertise scores of users as a mixture of gamma distributions.

\subsubsection{Summary}

We found a total of 18 studies across the selected portals focused on expert identification. Table~\ref{tab:methods_comparison} presents the summary of eight most cited and representative studies of the corresponding methods across four Q\&A platforms, namely, Quora, Reddit, Stack Overflow and Yahoo! Answers. We can notice that the maximum number of categories that the authors of these articles analysed is two -- expert and non-expert, omitting out-of-scope comments. In all of these cites except Yahoo! Answers, statistics and NLP were introduced; in turn, network analysis was applied to all except Quora. Furthermore, the studies describing all the mentioned portals used crowdsourced features, while they did not employ user features in Reddit. Finally, in half of the platforms, more specifically in Reddit and Yahoo! Answers, the authors did not characterise experts, neither they performed manual labelling by experts.

\begin{table*}[]
\resizebox{\textwidth}{!}{%
\begin{tabular}{|l|l|c|c|c|c|c|c|c|c|c|}
\hline
\textbf{\begin{tabular}[c]{@{}l@{}}Work performed\\ expert identification (EI)\end{tabular}} &
  \textbf{\begin{tabular}[c]{@{}l@{}}Q\&A \\ platform\end{tabular}} &
  \textbf{\begin{tabular}[c]{@{}c@{}}Number of \\ categories\end{tabular}} &
  \textbf{\begin{tabular}[c]{@{}c@{}}Manual labelling\\ by experts\end{tabular}} &
  \textbf{\begin{tabular}[c]{@{}c@{}}Statistics \\ for EI\end{tabular}} &
  \textbf{\begin{tabular}[c]{@{}c@{}}Network analysis \\ for EI\end{tabular}} &
  \textbf{\begin{tabular}[c]{@{}c@{}}NLP\\ for EI\end{tabular}} &
  \textbf{\begin{tabular}[c]{@{}c@{}}ML for\\ EI\end{tabular}} &
  \textbf{\begin{tabular}[c]{@{}c@{}}Crowdsourced\\ features\end{tabular}} &
  \textbf{\begin{tabular}[c]{@{}c@{}}User\\ features\end{tabular}} &
  \textbf{\begin{tabular}[c]{@{}c@{}}Characterising\\ experts\end{tabular}} \\ \hline
\citep{patil2016detecting} &
  Quora &
  2 &
  \yestick &
  \notick &
  \notick &
  \yestick &
  \yestick &
  \yestick &
  \yestick &
  \yestick \\ \hline
\citep{zhao2014expert} &
  Quora &
  2 &
  \notick &
  \yestick &
  \notick &
  \notick &
  \notick &
  \yestick &
  \yestick &
  \notick \\ \hline
\citep{choi2015characterizing} &
  Reddit &
  2 &
  \notick &
  \notick &
  \notick &
  \yestick &
  \notick &
  \yestick &
  \yestick &
  \notick \\ \hline
\citep{lim2017estimating} &
  Reddit &
  2 &
  \notick &
  \yestick &
  \notick &
  \notick &
  \notick &
  \yestick &
  \notick &
  \notick \\ \hline
\citep{sumanth2018discovering} &
  StackOverflow &
  2 &
  \notick &
  \notick &
  \yestick &
  \notick &
  \notick &
  \yestick &
  \notick &
  \notick \\ \hline
\citep{van2015early} &
  StackOverflow &
  2 &
  \notick &
  \notick &
  \notick &
  \yestick &
  \yestick &
  \yestick &
  \yestick &
  \notick \\ \hline
\citep{bouguessa2008identifying} &
  Yahoo! Answers &
  2 &
  \yestick &
  \notick &
  \yestick &
  \notick &
  \notick &
  \yestick &
  \notick &
  \notick \\ \hline
\citep{10.1145/1321440.1321575} &
  Yahoo! Answers &
  2 &
  \notick &
  \notick &
  \yestick &
  \notick &
  \notick &
  \yestick &
  \notick &
  \notick \\ \hline
\rowcolor[HTML]{9AFF99} 
\textbf{Our contribution} &
  Reddit &
  3 &
  \yestick &
  \notick &
  \notick &
  \yestick &
  \yestick &
  \yestick &
  \yestick &
  \yestick \\ \hline
\end{tabular}%
}
\caption{Comparison of existing methods across Q\&A platforms}
\label{tab:methods_comparison}
\end{table*}

The summary presented in \tableautorefname~\ref{tab:methods_comparison} also serves as a base to state the future contributions of the paper at hand. We selected Reddit as a Q\&A portal because there are already many studies focused on experts identification on Stack Overflow and Yahoo! Answers. Additionally, in our work, experts labelled the comments into three groups, namely, expert, non-expert and out-of-scope. Moreover, as far as we know, this is the first work that deployed NLP and ML methods using crowdsourced and user features for expert identification. We also characterised expert users which was rarely done before.

\section{Methodology}
\label{sec:methodology}

This section presents an overview of the whole methodological process pursued during our work. First, we explain the selection of the Q\&A portal by characterising its context and data access. Then, we describe the thread selection process that we followed. Further, we circumscribe the human-in-the-loop approach and, concretely, how experts manually labelled the comments from the selected Q\&A portal and the selected thread. Lastly, we explain the final data collection and present the details on how the ML model was used to identify experts in the field was built.

\subsection{Methodology overview}

\begin{figure*}[!ht]
\includegraphics[width=1\textwidth]{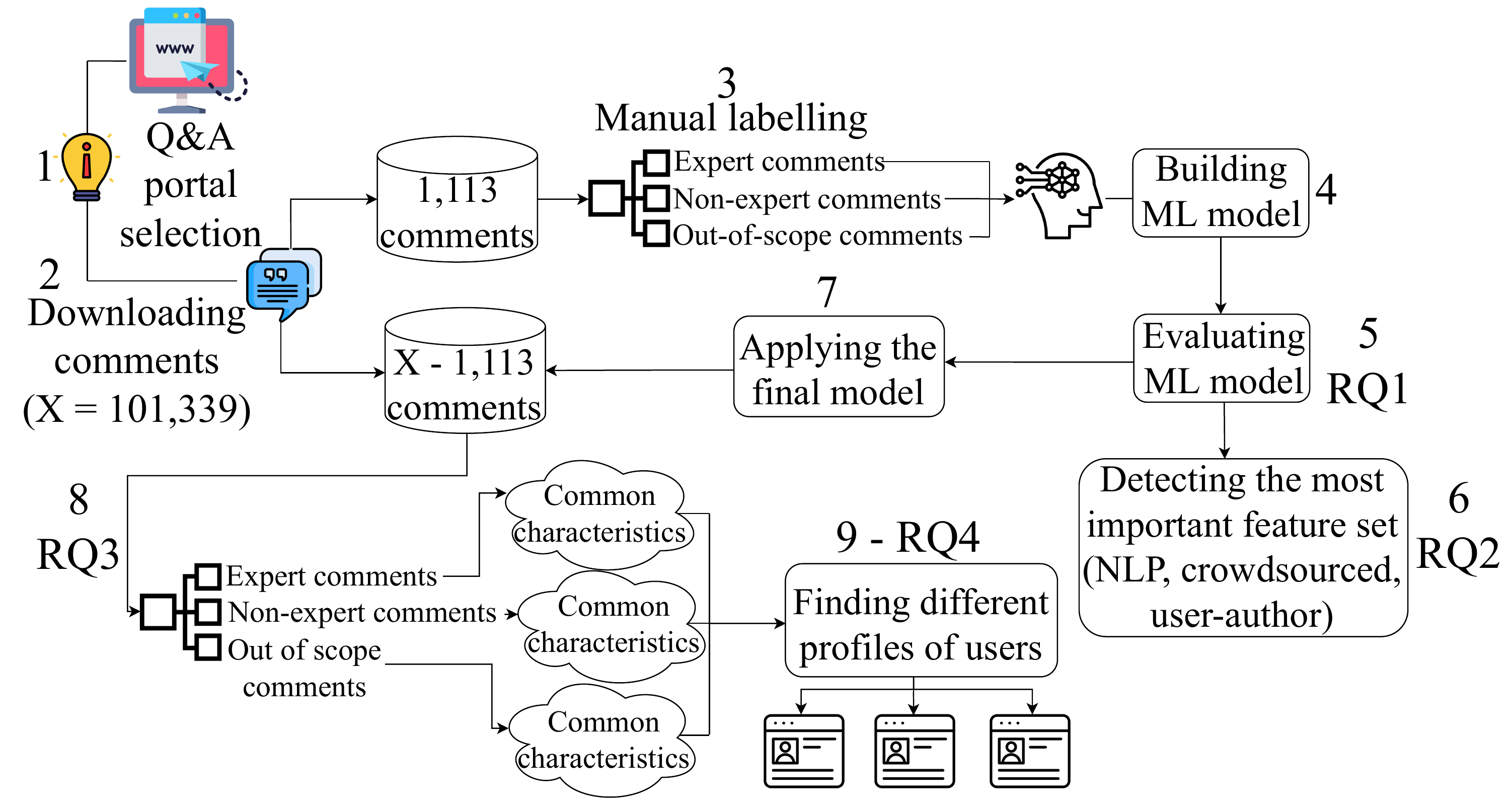}
\centering
\caption{Overview of the methodology to identify expert, non-expert and out-of-scope comments in Reddit}
\label{fig:methodology}
\end{figure*}

To answer the RQs stated in Section~\ref{sec:introduction}, we pursued the methodology process presented in Figure~\ref{fig:methodology}. In this figure, the numbers represent the sequential order of the steps as well as indicate each of the RQs.

In the first step, we selected the Q\&A portal and the thread for experts identification based on various metrics. In the second step, we downloaded all the posts and the respective comments and subcomments (comments to comments) of the selected thread in the selected portal from May 2020 to April 2021 because we consider it as a representative time span. Then, we randomly selected around 1,100 comments evenly distributed between 12 months in the timeline mentioned above from those posts that had between 5 and 20 comments (in this way, the post has a sufficient number of comments but still is not overloaded) without any additional particular criteria. In the third step, following the methodology described by Gobert et al.~\citep{gobert2013log}, two independent experts with relevant experience in the field manually coded the sample set following the previous criteria. In the fourth, we built the ML model to identify expert comments across the selected thread, and in the fifth step we evaluated the ML model. Next, in the sixth step, we detected the most important feature set among NLP, crowdsourced and user-author. In the seventh step, we applied the final model to the remaining comments. Then, the eighth step consisted in analysing these comments in order to obtain common characteristics of expert, non-expert and out-of-scope comments. Finally, in the ninth step, we used the final model to find different profiles of users in the selected portal.
%The model’s output was the comments distributed into three groups -- expert, non-expert and out-of-scope. 
%four experts in the field selected in Section~\ref{subsec:subreddit} analysed the comments of the selected thread and structured the emerged evidence into the above-described groups of criteria defining expert, non-expert and out-of-scope groups of comments.

\subsection{Q\&A portal selection}

Based on Table~\ref{tab:qa_comparison} presented in the previous section, we can conclude that the most active Q\&A portal is Reddit, being visited by 430 million active users every month. Moreover, as discussed earlier, it provides an API that can facilitate the step of downloading the required data. On the contrary, Reddit does not allow users who asked the questions to mark answers as correct. We believe that this is not an obstacle to conduct a study aiming to identify experts because, at any rate, these data could not serve as ground truth. Besides, based on Table~\ref{tab:methods_comparison}, we see not only that there is little research done on expert identification across Reddit, but also that the existing studies focused on this portal did not use ML and user features for expert identification. Ultimately, we believe that the absence of research manually labelling Reddit data by experts is an important gap, making it the right match for our study.

Reddit is one of the most used discussion websites, calling itself “home to thousands of communities, endless conversation, and authentic human connection.” It provides an opportunity to find communities of various interests for both personal and professional endeavours, which helps to maintain a diverse number of opinions, ideas and fellows.

\begin{figure}[!ht]
\includegraphics[width=0.5\textwidth]{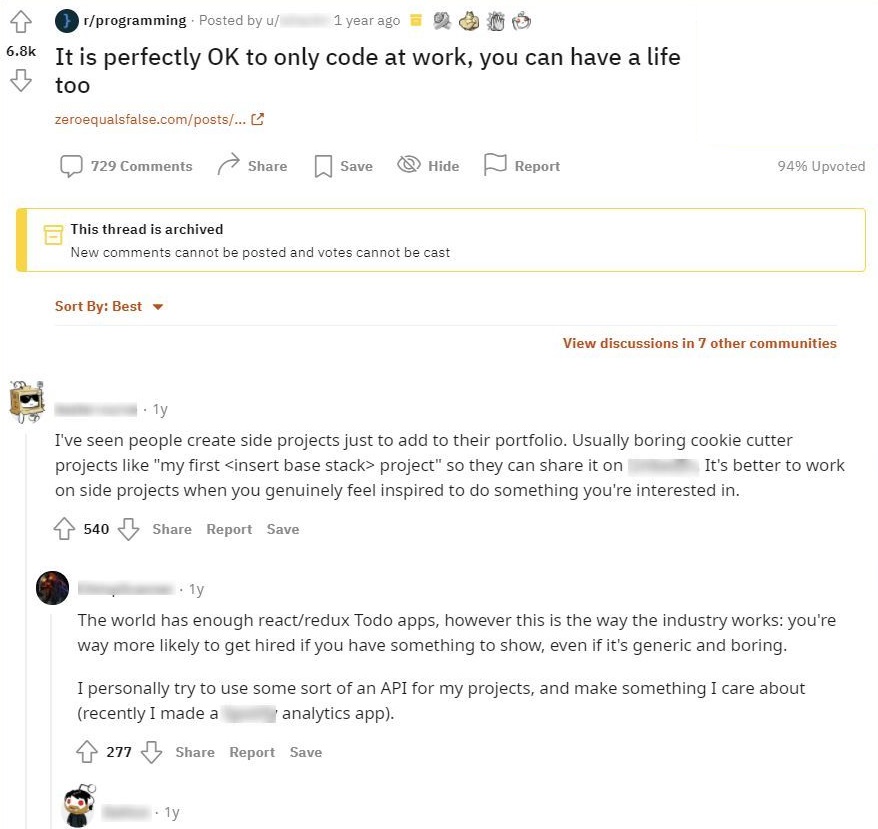}
\centering
\caption{Reddit website: post overview}
\label{fig:reddit_post}
\end{figure}

On Figure~\ref{fig:reddit_post}, we represent one particular post with more details which can be accessed by clicking on the direct link. Next to the post, we see its score, which is counted as the number of upvotes minus the number of downvotes, along with the arrows for upvoting or downvoting, the number of comments, the nickname of the author and their awards. We can observe the fact that users can comment not only on the initial posting but also on other comments (writing subcomments) and vote on the post or comments. Subcomments also can be upvoted and downvoted. A user may share, save, hide or report the post. According to Choi et al., each community on Reddit shows different characteristics; specifically, news-related, image-based, and discussion-related communities are more likely to have large, responsive, and viral conversations~\citep{choi2015characterizing}, which make this research more promising and exciting.

\subsection{Reddit API - data collection}

%Web data, particularly from APIs, have been a promising opportunity for researchers to use online social platforms’ databases of user-generated activity and content~\citep{baumgartner2020pushshift}. API is a programming tool with a set of functions provided by digital media companies that allow building and integrating applications’ software that interacts with their systems~\citep{hampton2017studying}.

The Reddit API allows users to post or extract data, users, variables, time periods, types of content, across concrete subreddits. It is advisable to download all the data at once to avoid the issue of inconsistency. However, depending on the search and unit of analysis, there are limits on the amount of data the system will return~\citep{amaya2021new}. Moreover, Reddit API caps the ability to pull data to 60 items (a single post or a comment along with the metadata associated with that submission) per minute. Thus, we decided to use Python Reddit API Wrapper (PRAW)\footnote {\url{https://github.com/praw-dev/praw/}} package to collect the data from several subreddits, which are described in the next section. The advantage of PRAW consists in the fact that it eases access to Reddit’s official API.

\subsection{Subreddit selection}
\label{subsec:subreddit}

In Reddit, all posts must be assigned to a subreddit, which can be created by any user and are also moderated by members of the Reddit community. Additionally, registered users can subscribe to the subreddits they want to follow, which makes the Reddit experience customisable. In this way, users will see only those posts relevant to their interests. Therefore, in our work, it was important to select the relevant subreddit not randomly but based on various metrics, which we represent in Table~\ref{tab:subreddit}.

Initially, we searched for subreddits related to the authors’ area of expertise of the paper at hand. Accordingly, we screened an unofficial list of subreddits\footnote{\url{https://www.reddit.com/r/ListOfSubreddits/wiki/listofsubreddits}} and then manually searched for additional ones that focus on data science, ML, artificial intelligence, big data and similar topics. In this way, 12 subreddits were singled out for a deeper screening. The most representative metrics for the final subreddit selection are unique users which identify the number of unique users in each subreddit, submission count and comment count representing the number of posts and comments, respectively, avg comments submissions -- an average number of comments and subcomments per post, avg comment length representing the average length of comments and subcomments, avg score and avg upvotes -- the average number of upvotes minus the number of downvotes and the number of only upvotes per post, avg total awards received displaying the average amount of awards received per post and avg submission upvote ratio showing the average upvote ratio per post.

\begin{table*}[]
\resizebox{\textwidth}{!}{%
\begin{tabular}{|l|l|l|l|l|l|l|l|l|l|}
\hline
\textbf{Subreddit} & \textbf{\begin{tabular}[c]{@{}l@{}}Unique\\ users\end{tabular}} & \textbf{\begin{tabular}[c]{@{}l@{}}Submission\\ count\end{tabular}} & \textbf{\begin{tabular}[c]{@{}l@{}}Comment\\ count\end{tabular}} & \textbf{\begin{tabular}[c]{@{}l@{}}Average comments\\ submissions\end{tabular}} & \textbf{\begin{tabular}[c]{@{}l@{}}Average comments\\ length\end{tabular}} & \textbf{\begin{tabular}[c]{@{}l@{}}Average\\ upvotes\end{tabular}} & \textbf{\begin{tabular}[c]{@{}l@{}}Average submission\\ upvote ratio\end{tabular}} & \textbf{\begin{tabular}[c]{@{}l@{}}Average total\\ awards received\end{tabular}} & \textbf{\begin{tabular}[c]{@{}l@{}}Average\\ score\end{tabular}} \\ \hline
\rowcolor[HTML]{67FD9A} 
datascience & 2,440 & 824 & 5,400 & 6.55 & 307.89 & 7.15 & 0.80 & 0.02 & 7.15 \\ \hline
MachineLearning & 2,419 & 1,525 & 5,071 & 3.33 & 359.47 & 5.51 & 0.89 & 0.02 & 5.51 \\ \hline
ArtificialInteligence & 505 & 1,575 & 507 & 0.32 & 261.36 & 2.03 & 0.98 & 0.01 & 2.03 \\ \hline
deeplearning & 386 & 268 & 502 & 1.87 & 259.70 & 3.29 & 0.75 & 0.02 & 3.29 \\ \hline
OpenAI & 47 & 14 & 81 & 5.79 & 156.80 & 2.95 & 0.85 & 0.02 & 2.95 \\ \hline
GPT3 & 202 & 65 & 569 & 8.75 & 228.76 & 3.56 & 0.88 & 0.03 & 3.56 \\ \hline
datamining & 12 & 10 & 4 & 0.40 & 304.25 & 3.36 & 0.97 & 0.07 & 3.36 \\ \hline
DeepLearningPapers & 25 & 28 & 23 & 0.82 & 231.00 & 4.29 & 0.81 & 0.06 & 4.29 \\ \hline
neuralnetworks & 95 & 76 & 106 & 1.39 & 185.47 & 3.84 & 0.81 & 0.01 & 3.84 \\ \hline
artificial & 523 & 349 & 776 & 2.22 & 203.36 & 5.68 & 0.81 & 0.02 & 5.68 \\ \hline
bigdata & 141 & 378 & 156 & 0.41 & 275.89 & 2.04 & 0.91 & 0.02 & 2.04 \\ \hline
deepmind & 0 & 2 & 0 & 0.00 & 0.00 & 11.00 & 0.87 & 0.00 & 11.00 \\ \hline
\end{tabular}%
}
\caption{Metrics for subreddit selection in Reddit}
\label{tab:subreddit}
\end{table*}

After analysing the metrics mentioned above, we selected the datascience subreddit\footnote{\url{https://www.reddit.com/r/datascience/}} as it proved to be the most active one in terms of the number of users and their activity. Moreover, with the increasing ability to collect, store and analyse an ever-growing diversity of data that are being generated with increasing frequency, the field of data science is growing significantly~\citep{saltz2015need}. We observed that there is much interest in this subreddit, so the users are involved in the discussions and are motivated to answer the questions well. Thus, we believe that the selected community provides an excellent opportunity for finding data science experts.

\subsection{Manual coding of data science expertise}
\label{subsec:manual_coding}

In this research we decided to use supervised learning approaches by performing human-in-the-loop manual coding of the ground truth toward creating a reliable algorithm~\citep{xin2018accelerating}. The rationale is that the original comments of Reddit are not selected as correct or not by other users. Therefore, the data sets need to be labelled given specific instructions. This process starts with experts investing their time in applying domain knowledge to label the data. There are several practical benefits of this approach. Firstly, tedious and mechanical iterations from active development time can be removed. Secondly, it brings an advantage of transparency since human intelligence is understandable and can be documented well. Lastly, it can incorporate human judgement into algorithms effectively. That being the case, human-in-the-loop design can improve the machine’s performance providing a balance between automated functionality and human interaction. Besides, experts can act in a straightforward way on simple tasks such as binary labels on objective tasks for deciding which label is correct when different annotators disagree. However, for subjective tasks, or even tasks with continuous data, no simple heuristic exists for deciding the correct solution~\citep{monarch2021human}. In this way, it is essential to build understandable criteria for labelling so that other researchers can re-use them.

In the scope of our work, we first analysed the thread of data science in Reddit and saw that there was evidence not only of expert comments relevant in the field of data science and non-expert comments not relevant in the field of data science, but also out-of-scope comments meaning that they did not correlate with the field of data science. Accordingly, instead of following the common approach of using two classes of data, we decided to include the out-of-scope group in order to improve the results of our work and attempt to differentiate better these classes of comments. In the next step, two experts read the comments to identify, name and categorise the type of the comments. First, both raters separately labelled 20 initial comments to understand the challenges of the process. It is essential to mention that for manual coding, we considered only comments to the initial posts, not taking into account subcomments because the latter did not provide much useful information, while we tested the ML model on both, comments and subcomments. To validate that a replicable construct was being coded, interrater reliability was established across two human coders~\citep{gobert2013log} which consisted in computing Cohen’s kappa. Next, the two coders discussed the coding scheme and the differences in their coding. Then, they coded the same 20 comments together. The following step consisted of separately coding an additional set of 20 comments. We again performed the inter-agreement validation by calculating Cohen’s kappa. This process repeated when Cohen’s kappa coefficient reached the established minimum equal to 0.7, which is a sufficient level of agreement better than chance~\citep{baker2008labeling}. Afterwards, the remaining comments were equally divided for each coder to code independently and discuss the difference in coding results to reach a consensus.

The authors of~\citep{das2015towards} defined \emph{data science} as \emph{``the use of statistical and ML techniques on big multi-structured data in a distributed computing environment to identify correlations and causal relationships, classify and predict events, identify patterns and anomalies, and infer probabilities interest and sentiment.''} [p. 2072] In the present case study, choosing correct criteria to identify experts is vital for the reason that a human coder needs sufficient information to label the data properly but should not be burdened with external information that could provoke a decline in identification speed and accuracy. The detailed criteria to identify expert, non-expert and out-of-scope comments in the data science field that we built is described in~\ref{sec:appendix1}. Since the data science field is vast, we performed a parameter analysis and decided to consider the posts that compiled at least three of the mentioned criteria as an expert response. While some of these criteria are distinctive, we acknowledge that there are literal overlaps among them because various subfields of data science are correlated. Also, based on the initial comments observation and review, if in one comment there was evidence of several classes, two data science experts chose the strongest following the schema presented next:

\begin{itemize}

    \item Expert AND Out-of-scope \textrightarrow Expert
    
    \item Non-expert AND Out-of-scope \textrightarrow Non-expert
    
    \item Expert AND Non-expert \textrightarrow Expert

\end{itemize}

\subsection{Description of the final data collection}

\begin{figure*}[!ht]
\includegraphics[width=1\textwidth]{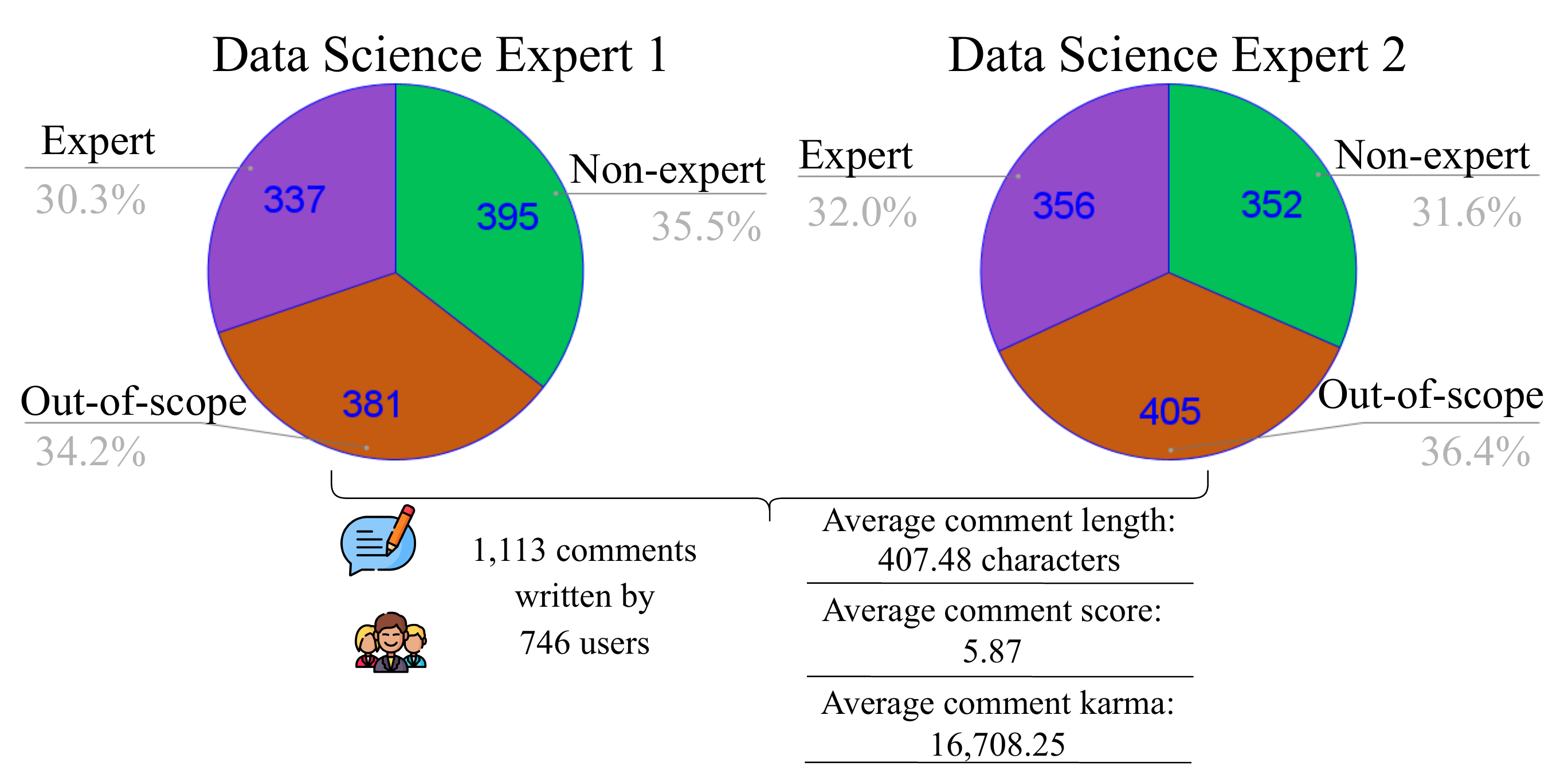}
\centering
\caption{Distribution of manual coding}
\label{fig:distribution}
\end{figure*}

The manual classification process per iteration with the corresponding number of comments, Cohen’s kappa and the required time, are described in Table~\ref{tab:manual_class}. No time restrictions were introduced for labelling. Finally, it took roughly in average 12 hours for the first and the second human coders to tag all all comments. The inter-agreement coefficient of the final set of 1,113 comments is 0.703, transferring the fact that the high agreement between the raters is not accidental, and the distribution of manual classification is summarised in Figure~\ref{fig:distribution}. As we can see, the distribution is smoothed over the categories, so our data set is balanced, which means that it prevented classification bias in the training set.

\begin{table}[]
\centering
\begin{tabular}{|l|l|l|l|}
\hline
\textbf{\begin{tabular}[c]{@{}l@{}}Nº of\\ iteration\end{tabular}} &
  \textbf{\begin{tabular}[c]{@{}l@{}}Nº of\\ comments\end{tabular}} &
  \textbf{\begin{tabular}[c]{@{}l@{}}Cohen's \\ kappa\end{tabular}} &
  \textbf{\begin{tabular}[c]{@{}l@{}}Time spent \\ on labelling\end{tabular}} \\ \hline
1.        & 26 comments    & 0.15  & No data     \\ \hline
2.        & 36 comments    & 0.42  & 30 minutes  \\ \hline
3.        & 20 comments    & 0.6   & 25 minutes  \\ \hline
4.        & 21 comments    & 0.56  & 30 minutes  \\ \hline
5.        & 61 comments    & 0.48  & 60 minutes  \\ \hline
6.        & 20 comments    & 0.83  & 10 minutes  \\ \hline
7.        & 20 comments    & 0.6   & 10 minutes  \\ \hline
8.        & 178 comments   & 0.63  & 80 minutes  \\ \hline
9.        & 201 comments   & 0.75  & 90 minutes  \\ \hline
10.       & 734 comments   & 0.72  & 360 minutes \\ \hline
\rowcolor[HTML]{67FD9A} 
Final set & 1,113 comments & 0.703 & 530 minutes \\ \hline
\end{tabular}
\caption{Manual classification process per iteration}
\label{tab:manual_class}
\end{table}

\subsection{ML model to identify data science experts}

In this section, we describe the process of building a supervised learning model for the final goal of data science experts identification. First, we describe the data preprocessing step, followed by the feature engineering process. Then, we explain the supervised learning models that we chose for the stated goal and evaluation metrics to estimate their performance.

\subsubsection{Data preprocessing}

Before building a ML model, we followed several data preprocessing steps. This is important because ML algorithms are not able to work on the raw text directly. They heavily rely on a pre-defined set of features from the training data to produce output for the test data. Therefore, it is needed to convert the text into a matrix of features (transforming raw data into an understandable format), which is done with the help of NLP, whose goal is to train computers to process and accordingly understand a large amount of human language data. Another challenge of real-world data is that often they are incomplete, inconsistent, and are likely to contain many errors. Moreover, texts from social media have several linguistic peculiarities that may influence the classification
performance~\citep{farzindar2015natural}. Next, we describe data preprocessing steps that we followed as a proven method of resolving such issues, helping to get better results through classification algorithms:

\begin{enumerate}

    \item Remove blank rows in data.
    
    \item Change all the text to lower case.
    
    \item Word tokenisation is a process of breaking a piece of text into words, phrases, symbols, or other meaningful elements called tokens.
    
    \item Remove stop words -- commonly used words (e.g. “the,” “a,” “an,” “in,” etc.) that a search engine was programmed to ignore when indexing entries for searching and when retrieving them as the result of a search query. Stop words usually refer to the most common words in a language, but there is no single universal list of them.
    
    \item Remove non-alpha text -- any character (punctuation, symbol, etc.) that is not a number or letter is non-alphanumeric. In this step, we also remove code snippets, emojis and hyperlinks.
    
    \item Word transformations, which could be word stemming or word lemmatisation. Word stemming analyses the meaning behind a word. A stemmer operates on a single word without knowledge of the context and therefore cannot discriminate between words that have different meanings depending on the part of speech. Word lemmatisation, similarly to word stemming, is the process of finding the normalised form of a word, but here, this process involves using the context in which the word is being used. For this work, we decided to use the term lemmatisation.

\end{enumerate}

Next, using the term TF-IDF measure, we evaluated how relevant words in a collection of documents are. We used this measure because another scoring, namely Bag of Words, just creates a set of vectors containing the count of word occurrences in the document, while TF-IDF contains information on the more important words and the less important ones as well. As said, the TF–IDF score increases proportionally to the frequency of a word appearing in the document and decreases with the number of documents in the corpus that contain the word. Accordingly, a high TF-IDF value is obtained by a term that has a high frequency in a document and a low document frequency in the corpus. The TF-IDF score of words that appear in almost all documents is close to 0.

\subsubsection{Feature engineering for the ML model}

The most important part of text classification is feature engineering: the process of creating features for a ML model from raw text data. We cast the task of early detection of topical expertise as a classification problem: to decide whether a comment is an expert comment by using evidence from the portal and user’s behaviour on Reddit. We grouped the features that we obtained into three families:

\begin{enumerate}
    
    \item NLP features. These features include those that can be extracted from words, sentences, and phrases. The full list of NLP features is described in~\ref{sec:appendix2}. For further understanding, it is important to mention that one of these features is the probability of the comment to be an expert comment computed based on the TF-IDF estimated by Support Vector Machine (SVM).
        
    \item Crowdsourced features. These involve information or opinions from a group of people who submit their views via the Reddit site. These features include the karma of the comment and its score (the number of upvotes minus the number of downvotes).  
    
    \item User-author features. These features aim to gauge the activity level of those users who wrote the initial comment on Reddit. Accordingly, for each user whose comments happened to be labelled, we computed the number of comments and the number of posts per user throughout the stated timeline, the average number of words in posted answers and in posted questions, an average score per user, the number of days as a member of Reddit and average response time of every user.
    
\end{enumerate}

\subsubsection{Supervised learning model}
\label{subsec:supervised_model}

We used Logistic Regression, Random Forest (RF), Decision Tree (Dtree) and RuleFit over the features mentioned in the previous section to find potential experts because all four algorithms are generally known to perform very well for supervised learning problems. Moreover, they can be considered as interpretable models meaning that we are able to analyse the results in detail. Taking into account the importance of the combination of the selected features for identifying experts, we did not select the Naive Bayes algorithm because it makes an assumption that all the variables in the data set are not correlated to each other. All candidate ML models were built using the entire training set and all the features. To find the best combination of hyperparameters, we performed grid search. We used 10-fold cross-validation permitting us having less bias towards overestimating the true expected error. We did it in order to construct the training data and run the bagging algorithm over the various models mentioned above. Cross-validation ensures that the models do not overfit the classification models on the training sets, and they report the factual generalisation accuracy.

The development process started with an initial workflow containing simple data preprocessing and modelling steps. Then, based on the analysis of the resulting model, we modified the workflow to improve performance, such as using several feature selection methods, adding/removing features manually, adding regularisation to the model, and changing the evaluation metrics. The methods that we tested to get the best accuracy, AUC (Area Under the Receiver Operating Characteristic Curve), MAE (Mean Absolute Error) and R2 score metrics are as follows: removing features with low variance, selecting the best features based on univariate statistical tests (removing all but the \textit{k} highest scoring features, removing all but a user-specified highest scoring percentage of features), selecting features by recursively considering smaller and smaller sets of features, Forward-Sequential Feature Selection (SFS) that iteratively finds the best new feature to add to the set of selected features. Often, many of such iterations took place between the conception and the deployment of the ML model, with the developer as an integral component~\citep{xin2018accelerating}.

We used accuracy, AUC score, MAE and R2 score to evaluate the classification performances of different combinations of features and algorithms. 

\section{Results}
\label{sec:results}

\subsection{Expert, non-expert and out-of-scope comments identification - RQ1}

Table~\ref{tab:results_models} shows the performances of the models in predicting potential expert, non-expert and out-of-scope comments. We report the performance of the models using standard measures: accuracy, AUC score, MAE and R2 score. We see that the model that performs the best is RF reaching an accuracy of 0.82, an AUC score of 0.93, a MAE OF 0.24 degrees and R2 score of 0.57. Additionally, in Table~\ref{tab:confusion_matrix} we present a confusion matrix for the RF model, showing the summary of correct and incorrect prediction per comment type. As it can be observed, the proportion of true positive instances prevails. Additionally, we see that the model confused the most between the non-expert and out-of-scope comments (0.15\% with respect to the total number of errors). This could have been expected since these two comment types have certain pattern similarity due to their lack of details, irrelevant or misleading information. The introduced results may indicate that our models can solve this task precisely. Although this outcome is already satisfactory for a multi-class classification problem, we explore further the results of combining the sets and applying several feature selection methods.

\begin{table}[]
%\resizebox{\columnwidth}{!}{%
\centering
\begin{tabular}{|l|l|
>{\columncolor[HTML]{67FD9A}}l |l|l|}
\hline
 & \begin{tabular}[c]{@{}l@{}}Logistic\\ Regression\end{tabular} & \begin{tabular}[c]{@{}l@{}}Random\\ Forest\end{tabular} & \begin{tabular}[c]{@{}l@{}}Decision\\ Tree\end{tabular} & RuleFit \\ \hline
Accuracy & 0.78 & \textbf{0.82} & 0.74 & 0.76 \\ \hline
AUC & 0.91 & \textbf{0.93} & 0.87 & 0.87 \\ \hline
\begin{tabular}[c]{@{}l@{}}Mean Absolute\\ Error (in degrees)\end{tabular} & 0.24 & \cellcolor[HTML]{67FD9A}\textbf{0.2} & 0.32 & 0.35 \\ \hline
R2 Score & 0.57 & \cellcolor[HTML]{67FD9A}\textbf{0.66} & 0.46 & 0.48 \\ \hline
\end{tabular}
%}
\caption{Results comparison of Logistic Regression, Random Forest, Decision Tree and RuleFit models by accuracy, AUC, Mean Absolute Errors and R2 Score metrics}
\label{tab:results_models}
\end{table}

\begin{table}[]
\centering
\begin{tabular}{|l|l|l|l|}
\hline
\textbf{\begin{tabular}[c]{@{}l@{}}Type of comment\end{tabular}} &
  \textbf{\begin{tabular}[c]{@{}l@{}}Expert\end{tabular}} &
  \textbf{\begin{tabular}[c]{@{}l@{}}Non-expert\end{tabular}} &
  \textbf{\begin{tabular}[c]{@{}l@{}}Out-of-scope\end{tabular}} \\ \hline
Expert        & 22.04\%    & 2.42\%  & 1.08\%     \\ \hline
Non-expert        & 6.18\%    & 31.18\%  & 7.8\%  \\ \hline
Out-of-scope        & 0.27\%    & 4.3\%   & 24.73\%  \\ \hline
\end{tabular}
\caption{Confusion matrix for the Random Forest classifier}
\label{tab:confusion_matrix}
\end{table}

\subsection{The most important feature set for comments type detection - RQ2}

The first step was to test the selected RF model separately on all feature sets formulated earlier (NLP, crowdsourced, and user-author). Then, to identify the most important feature set for detection of topical expertise by performing feature selection discussed in Section~\ref{subsec:supervised_model}. The results of RF in the selected features sets are represented in Table~\ref{tab:importance_features}. As it can be observed, among individual features sets, the NLP features set outperformed the two other sets. Besides, applying user and crowdsourced features sets emphasised the fact that still comments themselves played a more important role. We believe that some of the features belonging to this set in fact can reflect the expertise of users because more experienced users tend to write more elaborated answers, which are harder to understand. Surprisingly, the use of the user features along with crowdsourced features did not improve the performance of the model. Finally, after applying feature selection methods, the combination of NLP and user sets helped us to obtain a slightly better result peaking the AUC score at 0.93, accuracy at 0.83, MAE at 0.15 degrees and R2 score at 0.69, which supports the strong predictive power of this optimal set of features.

\begin{table}[]
\resizebox{\columnwidth}{!}{%
\begin{tabular}{|l|c|c|c|l|c|c|c|l|
>{\columncolor[HTML]{67FD9A}}c |}
\hline
 & \multicolumn{1}{l|}{\textbf{User features}} & \multicolumn{1}{l|}{\textbf{Crowdsourced features}} & \multicolumn{1}{l|}{\textbf{NLP features}} &  & \multicolumn{1}{l|}{\textbf{\begin{tabular}[c]{@{}l@{}}Selected User \\ + NLP features\end{tabular}}} & \multicolumn{1}{l|}{\textbf{\begin{tabular}[c]{@{}l@{}}Selected Crowdsourced +\\ NLP features\end{tabular}}} & \multicolumn{1}{l|}{\textbf{\begin{tabular}[c]{@{}l@{}}Selected User +\\ Crowdsourced features\end{tabular}}} &  & \multicolumn{1}{l|}{\cellcolor[HTML]{67FD9A}\textbf{\begin{tabular}[c]{@{}l@{}}Selected all\\ features\end{tabular}}} \\ \cline{1-4} \cline{6-8} \cline{10-10} 
\textbf{Accuracy} & 0.51 & 0.44 & 0.82 &  & 0.82 & 0.82 & 0.51 &  & 0.83 \\ \cline{1-4} \cline{6-8} \cline{10-10} 
\textbf{AUC} & 0.7 & 0.61 & 0.93 &  & 0.93 & 0.93 & 0.7 &  & 0.93 \\ \cline{1-4} \cline{6-8} \cline{10-10} 
\textbf{\begin{tabular}[c]{@{}l@{}}Mean Absolute\\ Error (in degrees)\end{tabular}} & 0.57 & 0.59 & 0.17 &  & 0.2 & 0.19 & 0.62 &  & 0.15 \\ \cline{1-4} \cline{6-8} \cline{10-10} 
\textbf{R2 Score} & 0.12 & 0.27 & 0.67 &  & 0.66 & 0.68 & 0.3 &  & 0.69 \\ \hline
\end{tabular}%
}
\caption{Feature importances comparison of Random Forest by accuracy, AUC, Mean Absolute Errors and R2 Score metrics}
\label{tab:importance_features}
\end{table}

For obtaining this result, the model used the following features with the corresponding variable importance: TF-IDF derived features (0.42), reading time (0.09), number of words (0.09), response time (0.08), the average number of words in comments per author (0.06), automated readability index (0.05), Flesch reading ease (0.05), the subjectivity of the comment (0.05), the number of programming terms (0.03), number of posts per user (0.03), Smog index showing how many years of education are needed to understand a piece of writing (0.03), the average number of words in posts per author (0.03). We can notice that these features are derived from the user activity on Reddit and from the comments themselves. This means that crowdsourced metrics such as the comment karma are not representative in distinguishing expert and non-expert comments. On the other hand, we expected to see that NLP features contribute to identifying topical expertise since existing studies explored earlier showed that this was a well-known approach. Moreover, user features represent how active users were historically on Reddit, advancing our method.

\subsection{Common characteristics of expert, non-expert and out-of-scope comments - RQ3}

\begin{figure*}[!ht]
\includegraphics[width=1\textwidth]{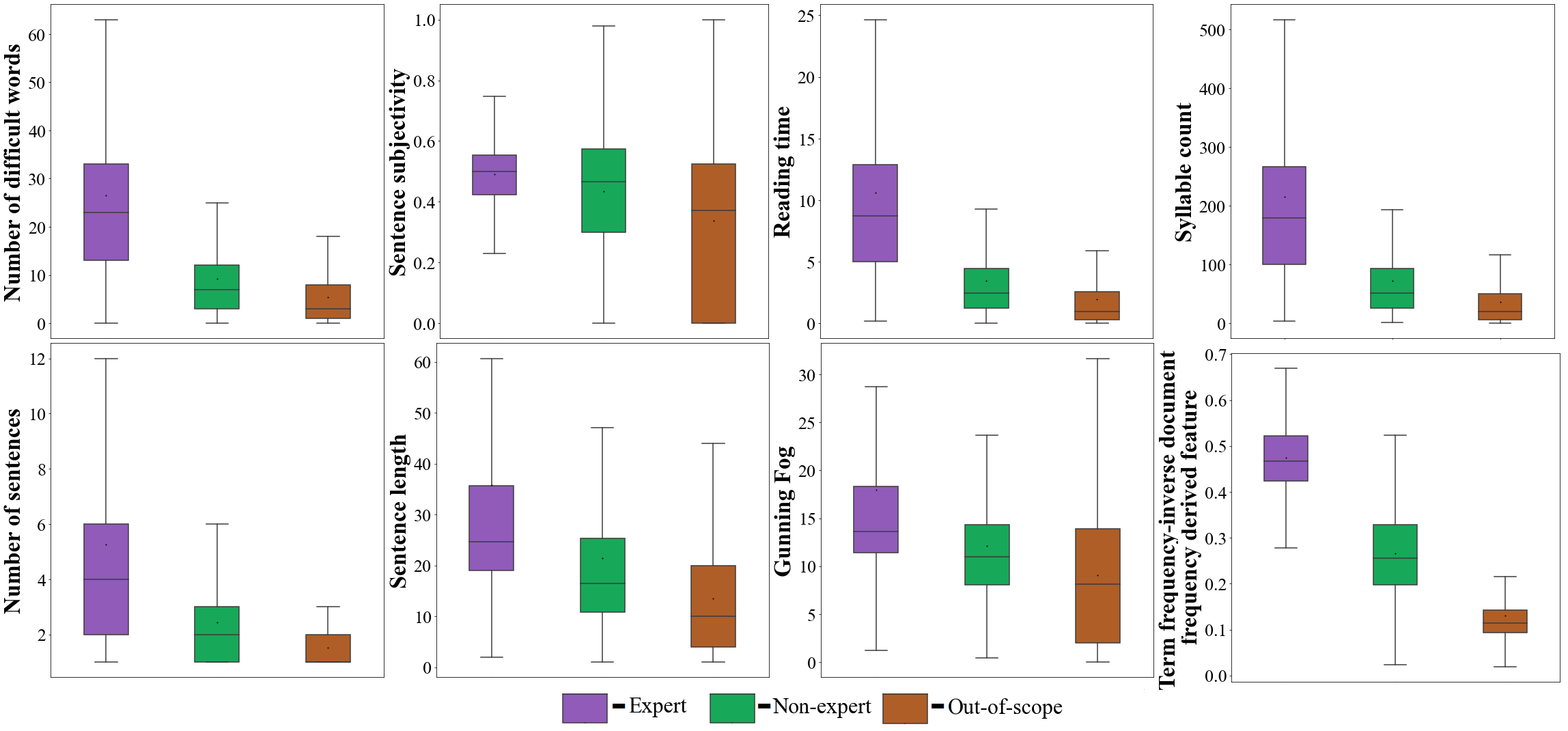}
\centering
\caption{Distribution of the most significant features by comment type (expert, non-expert and out-of-scope)}
\label{fig:characteristics_comments}
\end{figure*}

To answer this RQ, we predicted expert, non-expert and out-of-scope comments in the rest of the comments in our sample. The distribution resulted in having 13,798 expert, 57,896 non-expert and 28,328 out-of-scope comments. Two data science experts who performed the manual coding of comments described in Section~\ref{subsec:manual_coding} manually checked a random sample of 1,000 comments and agreed on 87\% of the predictions made by the model. We also performed multivariate analysis of variance (MANOVA) to ascertain that the differences between these types of comments are statistically significant. This fact was confirmed by obtaining an $F$-value = 8,583 and $p$-value $ \ll $ 0.0. Therefore, we can confirm that each type of comment has statistically significant different characteristics. Next, we conducted analysis of variance (ANOVA) for each individual feature to see which of them are statistically different. Accordingly, the examination of the corresponding common features of the three groups of comments is represented in Figure~\ref{fig:characteristics_comments}.

Based on this figure, we noticed several promising patterns in the out-of-scope comments, which can be helpful to identify these, serving to the early detection of spam or malicious messages. Firstly, the average number of difficult words of out-of-scope comments is much lower (5) than of expert (27) and non-expert (9) ones, which again could emphasise the fact of low-quality of their content. It is curious that the subjectivity of out-of-scope comments varies from 0 to 0.7 meaning that the amount of personal opinion and factual information contained in the text varies significantly what could serve as a base to distinguish between generic spam messages and more personalised out-of-scope comments. Finally, the average time required for reading out-of-scope comments is 2 seconds, while to read the non-expert and expert comments, a person needs 3 and 11 seconds, respectively. This is correlated with the low average number of syllabic words (36), whilst expert and non-expert comments have, on average, 215 and 71 syllabic words. 

With respect to expert comments, there are also encouraging characteristics. For example, on average, expert comments consist of more sentences equal to 5 with the average sentence length reaching 36 symbols; in contrast, non-expert and out-of-scope comments both have 2 sentences with a shorter length on average. Several readability scores show outstanding results for expert comments. These include, for example, average Gunning Fog representing the estimation of the years of formal education a person needs to understand the text on the first reading (18 for expert, 12 and 9 for non-expert and out-of-scope comments). Accordingly, the higher scores indicate text that is easier to read, and lower numbers mark text that is more difficult to read. This means that expert comments tend to provide more complicated content related to the data science topic. Finally, it is important to mention that the probability of the comment being an expert comment computed based on the TF-IDF estimated by SVM is also a representing feature (0.47 for expert comments, 0.27 and 0.13 for non-expert and out-of-scope comments), highlighting again the fact that the comment itself is prevalent.

\subsection{Existing profiles of Reddit users in terms of expertise - RQ4}

To answer this RQ, we aggregated the results by user. First, from the authors of comments identified as expert, non-expert and out-of-scope in the previous steps, we selected only those ones who wrote at least five comments and posts. By applying this measure, we make sure that we analyse only those users who showed a minimum activity in Reddit during the last year. As a result of this step, we selected 66,948 comments (9,874 expert, 39,698 non-expert and 17,376 out-of-scope comments) written by 3,945 unique users. For all unique users, we computed the percentage of expert, non-expert and out-of-scope comments that they wrote. Accordingly, we considered as expert users those who wrote at least 50\% of expert comments, non-experts those users who wrote at least 50\% of non-expert comments and out-of-scope users those who wrote at least 50\% of out-of-scope comments. We did not analyse those users that did not fall in any of the above-mentioned groups, because their profiles could not be interpreted distinctively. As a result, there are 79 experts, 2,495 non-experts and 210 out-of-scope users. It is noteworthy to mention that there are many more non-expert users than two other types of users. It is consistent with the distribution of comments presented in the previous section where we obtained more instances of non-expert comments. Moreover, it is not a trivial task to produce expert or out-of-scope comments in at least half of the comments.

\begin{figure}[h!]
\includegraphics[width=9.5cm, height=8cm]{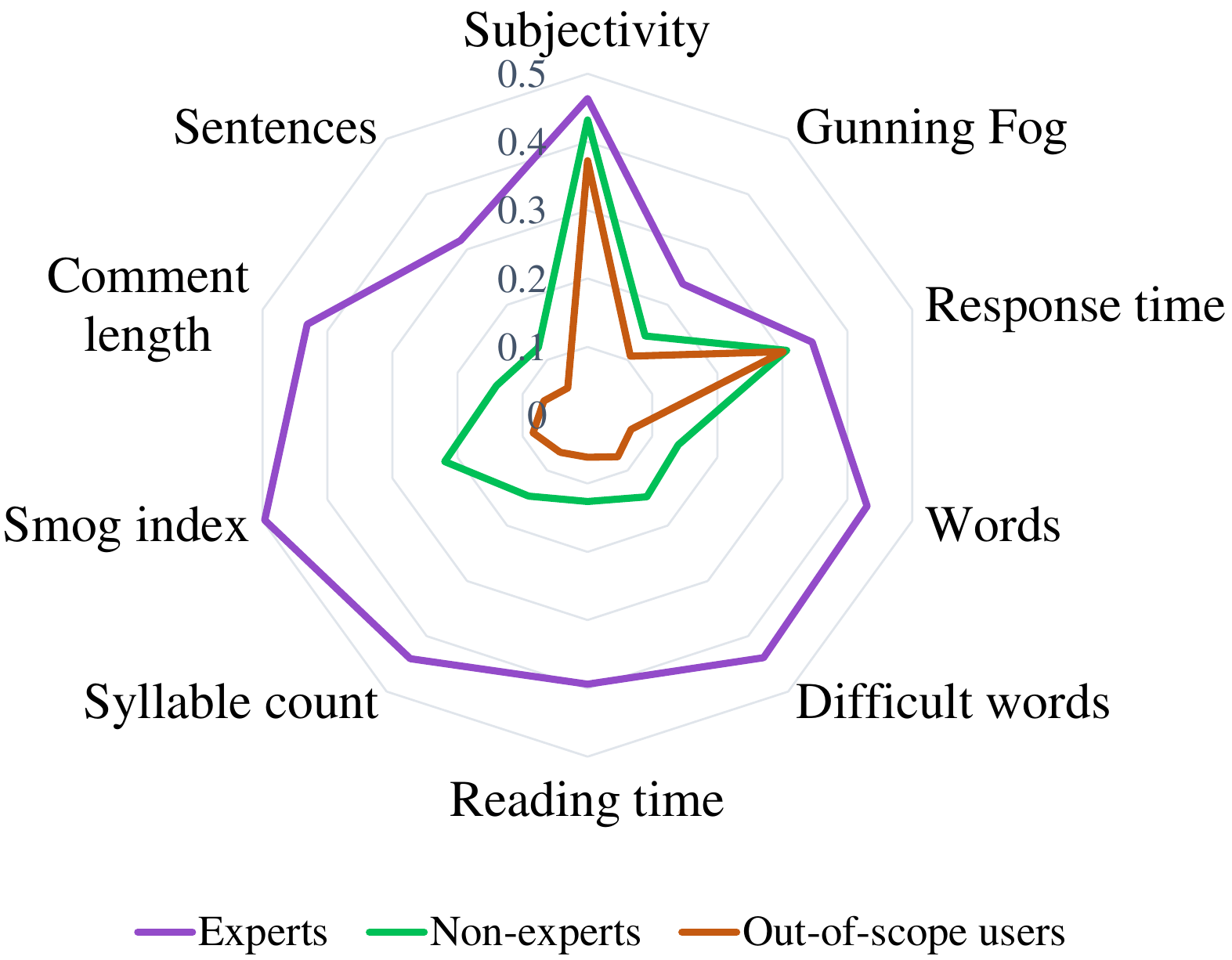}
\centering
\caption{Experts, non-experts and users writing out-of-scope comments}
\label{fig:radar_chart}
\end{figure}

In Figure~\ref{fig:radar_chart}, we represent these three types of users, from which it can be observed that experts differ from the two other types of users. Firstly, they are writing longer comments represented by a higher number of sentences, words and comment length. Furthermore, their comments consist of more difficult and syllable words. This causes an elevated reading time, Smog index and Gunning Fog, which represent how many years of education an average person needs to have to understand a comment. In opposition, the subjectivity of comments of all three types of users is almost indistinguishable. Finally, the response time of experts, non-experts and out-of-scope users are also similar.
 
\section{Discussion}
\label{sec:discussion}

In this section, we present a discussion following the obtained results. We also talk about the potential application of our work in real scenarios. Finally, we raise the limitations of our work.

\subsection{Obtained results}

During our work, we firstly proved that it is possible to develop models that can identify expert, non-expert and out-of-scope comments peaking the AUC score at 0.93, accuracy at 0.83, MAE at 0.15 degrees and R2 score at 0.69. Based on these results and the respective best-selected features, we predicted the type of 100,226 comments. Next, we discussed the most representative features of expert and out-of-scope comments. Finally, we analysed 3,945 users, grouped them into the groups of experts, non-experts and out-of-scope users, and highlighted their common characteristics. We can conclude that it is a feasible task to detect not only expert comments but also experts who tend to provide helpful content in the Reddit community and who are active thread contributors. Contrastingly, we have a much larger sample of non-expert comments and users, which require manual verification of the reasons why they did not fall into the expert group. At last, the characteristics for out-of-scope comments that we presented are representative and can clearly distinguish them from the rest of the comments. We believe that by answering the initially stated RQs, we build future work directions for recognising not only data science expert comments but also spammers and malicious users whose influence is enormous nowadays. 

We also would like to outline the contributions of our work which are based on the state of the art expert identification methods presented in Table~\ref{tab:methods_comparison}. Firstly, as far as we are aware, this is the first time that expert identification in such an active Q\&A platform as Reddit was done with manual labelling of comments by experts. We have not found any work applying supervised learning model and user features for addressing expert identification problem. We are of the opinion that it is a significant novelty since they provide an additional source for making better predictions. Finally, we did not find any characterisation of expert users in the literature and by doing it, we not only filled this gap but also facilitated several important applications which are described next.

\subsection{Application in real scenarios}

Our final ML model that can identify experts in the data science field can have various applications since several possible domains can benefit from such a study. First of all, we can follow the work of Yan et al.~\citep{yan2019social} that developed a framework of collecting validations for members’ skill expertise in the LinkedIn\footnote{\url{https://www.linkedin.com/}} professional social network. This work proved the importance of estimating the users’ skill expertise at a large scale which, respectively, can serve as a base for predicting who is/can be hired for a job requiring a particular skill. Reddit also has become an emerging resource for talent recognition in recent years. In this way, the results of our work can be used to fill the gap between recruiters and candidates so that recruiters can find relevant candidates that fulfil the job description.

Secondly, it would be helpful if the Reddit community could display the expertise of every user next to the nickname. In this way, even the controversial posts and comments would be evaluated beforehand so the readers can rely more on the users who have already proved to be trustworthy. Moreover, it can be possible to rank expert users of each subreddit.

Thirdly, we expect it to be easy to extrapolate our work, focused on data science authorities, to a different topic. We are of the opinion that all user and crowdsourced, and many of NLP features can be re-used to identify experts in other fields. However, it would be useful to operate with topic-related dictionaries to count features related to the lexicon count.

Moreover, the analysis of the common characteristics of expert comments and users can serve as a suggestion for teachers and professors of all levels to get new ideas about what skills are to be developed and practised for the successful career prospects of their students. In a typical Q\&A community, every question has one or more tags indicating the required skills to answer this question. Correspondingly, these tags can be considered as skill areas that professors are interested in.

Another application is the recommendation system, which provides personalisation mechanisms by suggesting helpful answers to a questioner and interesting questions to a potential respondent. For instance, a user could be notified about new questions that are pertinent to his/her interests and expertise. This could minimise the percentage of the questions that are poorly or not answered at all. Furthermore, we have seen that we can build user reputation schemes that can capture a user’s impact and significance in the system. Such schemes could help provide the right incentives to users to be fruitfully and meaningfully active, reducing noise and low-quality questions and answers. Searching and ranking answered questions based on reputation/quality or user interests would help to find more easily answers and avoid posting similar questions.

Finally, by examining the prevalent traits in the out-of-scope group of comments, we can identify potentially malicious and unreliable users or social bots in the early stage and reduce their influence. As stated by Parra-Arnau et al., even though social networks provide an easy and immediate way of communication, there exist significant privacy threats provoked by inexperienced or even irresponsible users recklessly publishing sensitive material~\citep{parra2017shall}. For example, Pastor-Galindo et al.~\citep{pastor2020spotting} analysed the presence and behaviour of social bots in Twitter in the context of the Spanish general election. The authors classified users as social bots or humans, concluding that a non-negligible amount of bots actively participated in the election. This, in return, could affect the belief of the social media users while deciding whom to vote.

\subsection{Limitations}

However, our work has some limitations that we would like to acknowledge. First of all, the scope of our work was restricted to the identification of experts only in one subreddit. Respectively, the obtained results are limited to the data science subreddit that we chose based on the selected metrics. Secondly, despite the fact that two raters who classified the comments have experience and education related to the data science field and that the obtained Cohen’s kappa agreement proved to be high, the chance of the human factor could not be excluded, which is a natural consequence arising from a discontinuity between human capabilities and system demands~\citep{bevilacqua2018human}. Moreover, there are threats to validity because it is a challenging task to develop a perfect coding schema with no overlaps among the categories. Moreover, we labelled 1,113 comments out of 101,339 (1.1\%). Thus, despite being sure that our coding schema produced reliable results, further studies are required to confirm and generalise the coding process. Finally, while Reddit is a large and, moreover, the biggest Q\&A site, it would be useful to repeat the study with other portals. It would be valuable because these sites should, preferably, span users with more backgrounds and interests that differ from those of Reddit users. 

\section{Conclusions and Future Work}
\label{sec:conclusion}

Understanding the user base in a community-driven service is essential both from a social and a research perspective. We addressed the task of detection of topical expertise in the data science thread in Reddit. We proposed a robust way to define expertise based on the manual coding results where two data science experts labelled expert, non-expert and out-of-scope comments. We presented a semi-supervised approach using the activity behaviour of every user, including NLP, crowdsourced and user feature sets and demonstrated the effectiveness of our method. Our results proved that it is feasible to accurately predict whether it is an expert, non-expert or out-of-scope comment. Although the features to be used may vary, we concluded that for the data science thread on Reddit, the NLP and user features contribute the most to the better identification of these three classes. Therefore, we expect this method to generalise well within various applications.

Our future work will focus on the model generalisation to detect experts in other Q\&A portals, which represent discrepant knowledge domains and, therefore, skills of another nature. Moreover, we will aim to perform a case study following the real scenarios that we suggested earlier such as developing a recommendation system or identifying potentially unreliable users. We also plan to evaluate our method on other corpora as well as extend our features to capture more aspects of the topic expertise. In this way, it would be possible to track how a user's expertise evolves from one topic to another over time.

\section*{Acknowledgments}
This study was partially funded by the COBRA project (10032/20/0035/00), granted by the Spanish Ministry of Defence and by the SCORPION project (21661-PDC-21), granted by the Seneca Foundation of the Region of Murcia, Spain.

%Bibliography
\bibliographystyle{unsrtnat}  
\bibliography{references}

\begin{appendices}

\section{The criteria to identify expert, non-expert and out-of-scope comments}  
\label{sec:appendix1}

The criteria to be determinant to identify expert comments in the data science field are:

\begin{enumerate}
    
    \item Prior experience of the user in the data science field stated in the profile.
    
    \item Employment of the user in the field related to data science.
    
    \item A historical record of the user in Reddit, including awards that are considered as a way to recognise and react to each other’s contributions and karma -- a reflection of how much the users’ contributions mean to the community.
    
    \item Evidence of a user providing relevant and detailed feedback to the original question.
    
    \item Expression of information through mathematical formulae or programming language in a concise way.
    
    \item Clear evidence of data science expertise, including:
    
    \begin{itemize}
    
        \item Showing machine learning skills whose necessity is explained by the fact that companies navigate the data deluge and try to build automated decision systems that hinge on predictive accuracy~\citep{provost2013data}.
        
        \item Demonstrating the knowledge of statistics, especially Bayesian statistics, which indicates a working knowledge of probability, distributions, hypothesis testing, and multivariate analysis~\citep{dhar2013data}.
        
        \item Correct analysis of the heterogeneous and unstructured data, which requires integration, interpretation, and sense-making that is increasingly derived through tools from computer science, linguistics, econometrics, sociology, and other disciplines~\citep{dhar2013data}.
    
    \end{itemize}

    \item Crowdsourced metric of expertise obtained by counting the upvote ratio of the comment.
    
    \item Delay in reply, meaning that the response occurred sometime after the initial post was published.
    
    \item Novelty of the proposed answer obtained by a thorough search of the existing solutions.
    
    \item Proof of the ability to formulate problems in a way that results in practical solutions since it involves the ability to see commonalities across very different problems~\citep{dhar2013data}. It can be equivalent to the term of computational thinking described by Wing as the process that involves solving problems, designing systems, and understanding human behaviour by drawing on the concepts fundamental to computer science~\citep{wing2006computational}.
    
    \item Advising a tool with a detailed explanation.
    
    \item Generalise and transfer this problem-solving process to be able to solve a wide variety of problem families~\citep{plaza2021promoting}

\end{enumerate}

Non-expert responses are those comments which:

\begin{enumerate}

    \item Do not provide a considerable amount of details.
    
    \item Clearly state the absence of data science expertise.
    
    \item Crowdsourcing metric of expertise obtained by counting upvote ratio of the comment.
    
    \item Coding bugs, misuse of statistics, misleading interpretation or communication of the results meaning an unsatisfactory quality within data-focused projects~\citep{saltz2015need}. 
    
    \item Advising a tool without even a basic explanation.
    
\end{enumerate}

Out-of-scope comments include:

\begin{enumerate}

    \item Irrelevant or unsolicited messages (spam messages). Humour displays.
    
    \item Answers to polar or general questions whose expected answer is one of two choices, one that affirms and another that denies the question.
    
    \item Related but not detailed or not related follow-up questions.
    
    \item A positive or negative experience that is associated with a particular pattern of physiological activity.
    
    \item Expressions of a positive or negative personal experience of the user.
    
    \item Expressions of gratitude.
    
    \item Very short answers.
    
    \item References/links.
    
    \item Comments written in other languages apart of English.

\end{enumerate}

\section{The NLP feature set description}
\label{sec:appendix2}

The NLP feature set includes the following features:

\begin{enumerate}

\item Word count.

\item Syllable and polysyllable words count.

\item Comment length represented by the character count.

\item Average word length.

\item Average sentence length.

\item Sentiment analysis consisting in computing polarity and subjectivity (the average number of subjective words in posted answers) of sentences with TextBlob -- a Python library for processing textual data~\citep{loria2018textblob}.

\item Data science score calculated based on the count of data science terms that belong to the semantic word lists in the text snippet.

\item Analysis of text across counting words in several lexical categories (programming, technology) through a tool called Empath~\citep{fast2016empath} that can generate and validate new lexical categories on demand from a small set of seed terms. This feature represents a normalised value over words in each comment for each category.

\item The entropy of the answer which is a statistical parameter that measures how much information is produced on average for each letter of a text in a language.

\item The readability of the answers, including several metrics indicating how difficult a passage in English is to understand, such as the number of difficult words, reading time, the Flesch–Kincaid readability test, automated readability index, Flesch reading ease, the Spache readability formula, Coleman–Liau index, Dale–Chall readability score, Simple Measure of Gobbledygook (SMOG) and the Gunning fog index measuring how many years of education the average person needs to have to understand the text, amongst others. Complete definitions of these scores can be found in the Textstat Python library\footnote{\url{https://pypi.org/project/textstat/}}.

\item The probability of the comment to be an expert comment is computed based on the TF-IDF estimated by Support Vector Machine (SVM).

\end{enumerate}

\end{appendices}

\end{document}